# Causal Bounds and Instruments

**Roland R. Ramsahai**
Department of Statistics
University of Oxford
ramsahai@stats.ox.ac.uk

## Abstract

Instrumental variables have proven useful, in particular within the social sciences and economics, for making inference about the causal effect of a random variable, $B$, on another random variable, $C$, in the presence of unobserved confounders. In the case where relationships are linear, causal effects can be identified exactly from studying the regression of $C$ on $A$ and the regression of $B$ on $A$, where $A$ is the instrument. In the more general case, bounds have been developed in the literature for the causal effect of $B$ on $C$, given observational data on the joint distribution of $C$, $B$ and $A$. Using an approach based on the analysis of convex polytopes, we develop bounds for the same causal effect when given data on $(C,A)$ and $(B,A)$ only. The bounds developed are thus in direct analogy to the standard use of instruments in econometrics, but we make no assumption of linearity. Use of the bounds is illustrated for experiments with partial compliance. The bounds are, for example, relevant in genetic epidemiology, where the 'Mendelian instrument' $S$ represents a genotype, and where joint data on all of $C$, $B$ and $A$ may rarely be available but studies involving pairs of these may be abundant. Other examples of bounding causal effects are considered to show that the method applies to DAGs in general, subject to certain conditions.

## 1 Introduction

Studies in almost any physical, life and social science require inference on the causal effect of one variable on another. The causal effect is the expected change in one variable when we intervene and change the value of another variable. Such effects can sometimes be estimated from data obtained by the observation of changes in variables but can always be estimated when we have data obtained by setting certain variables. Cases where such estimation is not possible arise because of the presence of latent confounders, unobserved variables which affect more than one variable. If intervention data is not available then it is up to the statistician to provide the best possible inference from the given data. One approach which has proven quite useful in the social sciences and economics is to assume *linearity* and *additivity* and include instrumental variables (IV) in the analysis. Let $A$ be an instrumental variable for the effect of a random variable, $B$, on another random variable, $C$, in the presence of an unobserved confounder, $U$. This means that $A \perp\!\!\!\perp U$, $A \not\!\perp\!\!\!\perp B$ and $C \perp\!\!\!\perp A \mid (B,U)$. The graphical model (Lauritzen, 1996) in Fig.1 represents the relations $U \perp\!\!\!\perp A$ and $C \perp\!\!\!\perp A \mid (B,U)$. Assuming *stability* (Pearl, 2000), it represents the condition that $A$ is an instrument. Under the assumption of linearity and additivity between $C$ and $B$ and $U$, the causal effect of $B$ on $C$ can be identified exactly from studying the regression of $C$ on $A$ and the regression of $B$ on $A$, where $A$ is the instrument. The IV estimator of the causal effect of a unit change in $B$ on $C$ is $\hat{\beta}_{C\|B}$ and is given by

$$\hat{\beta}_{C\|B} = \frac{\hat{\beta}_{C|A}}{\hat{\beta}_{B|A}},$$

where $\hat{\beta}_{C|A}$ and $\hat{\beta}_{B|A}$ are estimates of the regression coefficients for the regression of $C$ on $A$ and $B$ on $A$ respectively (Durbin, 1954). An important characteristic of the IV estimator is that it can still be calculated if there are independent studies collecting data on the relationship of $(C,A)$ and $(B,A)$ only but not measuring all three. This approach may be deemed as not applicable for certain circumstances, as the assumption of linearity is sometimes far-fetched, and a non-parametric method would be more appropriate. In the literature, non-parametric bounds have been developed for causal effects in terms of observational data. Robins (1989) and Manski (1990) derive such bounds but, as Balke and Pearl (1997) point out, they are not the sharpest possible. Balke and Pearl (1997) formulates the problem as a linear programming problem in a counterfactual framework involving instrumental variables and develops causal bounds for the effect of $B$ on $C$, given observational data on the joint distribution of $C$, $B$ and $A$. They also show that their bounds are the sharpest possible. However, Dawid (2003) derives the same bounds in a probabilistic framework using the conditional independence relations represented in Fig.1 and the augmented DAG (Dawid, 2002). Both analyses focus on studies with imperfect compliance, where $A$ is treatment assignment, $B$ treatment received and $C$ the response. Pearl (1995) derives a falsifiable condition for $A$ to be an instrumental variable for the causal effect of $B$ on $C$ or for Fig.1 to hold. It is the 'instrumental inequality'

$$\max_b \sum_c \left[ \max_a P(c, b \,|\, a) \right] \leq 1, \qquad (1)$$

which represents a set of constraints on the joint distribution of $C$, $B$ and $A$. If the instrumental inequality is violated then Fig.1 is not a valid model for the data and none of the bounds in Balke and Pearl (1997) or Dawid (2003) hold. Therefore when trivariate data is available we can use the instrumental inequality as a test to determine whether Fig.1 is invalid and if it is not invalid we can bound the causal effect of $B$ on $C$. However, problems arise when we only have bivariate data, which is often the case in 'Mendelian randomisation' in genetic epidemiology (Didelez and Sheehan, 2007). In Mendelian randomisation, $A$ is the genotype, $B$ the phenotype and $C$ is the occurrence of a disease of interest. It is often the case that only genotype-phenotype and genotype-disease data are available. We therefore derive bounds on the causal effect of $B$ on $C$ when given $(C \,|\, A)$ and $(B \,|\, A)$ data and constraints which must be satisfied by the bivariate data if the model is valid. This is an important example as it is in direct analogy to the instrumental variable approach, only involving data on $(C \,|\, A)$ and $(B \,|\, A)$. Below we derive such bounds based on the methodology described in Dawid (2003). We also describe the general method of bounding distributions and causal effects in a DAG by simply exploiting the conditional independence relations between variables. Conditional independence restrictions from a DAG are simply expressed in terms of $P(C \,|\, B, A, U)$ and $P(B \,|\, A, U)$, whereas their observational consequences are represented in terms of $P(C, B, A \,|\, U)$. Application of the method is demonstrated with various other examples.

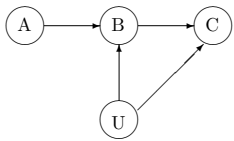

Figure 1: Instrumental variable

## 2 Properties of a Probability Distribution

Consider 4 random variables $A$, $B$, $C$ with sample spaces $\{1, 2\}$, $\{0, 1\}$, $\{0, 1\}$ respectively and $U$ with an unknown sample space. The probability distribution of $(C, B, A \,|\, U)$ can be represented by a vector

$$\vec{v} = (\xi^*_{001}, \xi^*_{011}, \xi^*_{101}, \xi^*_{111}, \xi^*_{002}, \xi^*_{012}, \xi^*_{102}, \xi^*_{112}),$$

where $\xi^*_{cba} = P(C, B, A \,|\, U)$. The random variable $U$ can be treated as a parameter of the distribution since $\vec{v}$ varies as $U$ varies. Without any assumptions we know that $\vec{v}$ is a point in a 7 dimensional subspace of $[0, 1]^8$ or a hyperplane in $[0, 1]^8$ since

$$\sum_c \sum_b \sum_a \xi^*_{c,b,a} = 1. \qquad (2)$$

Let the hyperplane represented by Eq.(2) be $\mathcal{Y}$. Therefore, by the axioms of probability, $\vec{v} \in \mathcal{Y}$. In other words, $\mathcal{Y}$ is the set of all vectors in $[0, 1]^8$ that represent probability distributions. From the factorisation of any joint probability distribution,

$$P(C, B, A \,|\, U) = P(C \,|\, B, A, U) P(B \,|\, A, U) P(A \,|\, U).$$

Therefore the vector $\vec{\tau} = (\eta_{01}, \eta_{11}, \eta_{02}, \eta_{12}, \delta_1, \delta_2, \psi)$, $\vec{\tau} \in [0, 1]^7$, where $\eta_{ba} = P(C = 1 \,|\, B, A, U)$, $\delta_a = P(B = 1 \,|\, A, U)$ and $\psi = P(A = 2 \,|\, U)$, can also be used to represent any probability distribution. The mapping is possible since

$$\sum_c P(C \,|\, B, A, U) = \sum_b P(B \,|\, A, U) = \sum_a P(A \,|\, U) = 1,$$

for all $B$ and $A$. Both $[0, 1]^7$ and $\mathcal{Y}$ are 7 dimensional. There exists a mapping

$$\Xi : \vec{\tau} \in [0, 1]^7 \to \vec{v} \in \mathcal{Y}.$$

For the complete graph, i.e. without any conditional independence relations

$$\Xi([0, 1]^7) = \mathcal{Y}.$$

## 3 Model

Consider a model in which $C \perp\!\!\!\perp A \,|\, (B, U)$, corresponds to the graphical model in Fig.2. For this model,

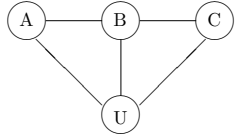

Figure 2: Model assuming $C \perp\!\!\!\perp A \,|\, (B, U)$

only values of $\vec{\tau}$ and $\vec{v}$ which exhibit the properties will fit the model. Consider a study in which $\vec{u}$ data is available, where

$$\vec{u} = (\xi_{001}, \xi_{011}, \xi_{101}, \xi_{111}, \xi_{002}, \xi_{012}, \xi_{102}, \xi_{112}).$$

and $\xi_{cba} = P(C, B, A)$. In order to derive falsifiable constraints for the model, we need to determine the set of possible $\vec{u}$. Since $\xi_{cba} = E_u[\xi^*_{cba}]$, then the set of possible $\vec{u}$ lies in the convex hull of the set of possible $\vec{v}$. However, the set of possible $\vec{v}$ for the model is not obvious but the set of possible $\vec{\tau}$ is simply the intersection of the hyperplanes

$$\eta_{01} = \eta_{02} \text{ and } \eta_{11} = \eta_{12}. \qquad (3)$$

Let $\mathcal{T}$ be the set of $\vec{\tau}$ which satisfy the model restrictions. Therefore $\mathcal{T}$ is the set of $\vec{\tau}$ which satisfy Eq.(3), $\mathcal{T} \subseteq [0, 1]^7$, $\dim(\mathcal{T}) = 5$ and $\Xi(\mathcal{T}) = \mathcal{V} \subseteq \mathcal{Y}$, where $\mathcal{V}$ is the set of possible $\vec{v}$ for the model. We can find the set of possible $\vec{v}$ by transforming $\mathcal{T}$. We only need 5 independent components of $\vec{\tau}$ to transform $\mathcal{T}$. Consider first the transformation of the extreme vertices of $\mathcal{T}$, $\hat{\mathcal{T}}$. Define $\hat{\mathcal{V}} := \Xi(\hat{\mathcal{T}})$. The extreme vertices are listed and the transformation is represented in Fig.3 where $\eta_0 = \eta_{01} = \eta_{02}$ and $\eta_1 = \eta_{11} = \eta_{12}$. Let $\mathcal{H}$ and $\hat{\mathcal{H}}$ be the convex hull of $\mathcal{V}$ and $\hat{\mathcal{V}}$ respectively. Since $\mathcal{H}$ is the convex hull of the set of possible $\vec{v}$ then it is the set of possible $\vec{u}$. We use a program such as *Polymake* (Gawrilow and Joswig, 2004) to find the representation of $\hat{\mathcal{H}}$ in terms of its facets or inequalities. The extreme vertices are entered in the same format as in Fig.3, with each column corresponding to a coordinate. Polymake outputs inequalities in the same format, where each number corresponds to the coefficient of the component its column represents in the inequality. The vector $\vec{u}$ must satisfy these inequalities to fit the model, i.e. to satisfy $C \perp\!\!\!\perp A \,|\, (B, U)$, since $\mathcal{H} = \hat{\mathcal{H}}$ (proof in Appendix A). The inequalities for this case are all trivial since here $\hat{\mathcal{H}} = \mathcal{Y}$. Note that this does not necessarily imply that $\mathcal{V} = \mathcal{Y}$.

The method described here determines the constraints on $\xi_{cba}$ for $C \perp\!\!\!\perp A \,|\, (B, U)$ but can easily be adjusted to find the constraints that quantities such as $P(C \,|\, A)$, $P(B \,|\, A)$, $P(C \,|\, B)$ etc. must satisfy. This is done by transforming $\mathcal{T}$ to a vector of quantities related to these terms instead of a vector of $P(C, B, A \,|\, U)$.

| $\eta_0$ | $\eta_1$ | $\delta_1$ | $\delta_2$ | $\psi$ | $\xi^*_{001}$ | $\xi^*_{011}$ | $\xi^*_{101}$ | $\xi^*_{111}$ | $\xi^*_{002}$ | $\xi^*_{012}$ | $\xi^*_{102}$ | $\xi^*_{112}$ |
|---|---|---|---|---|---|---|---|---|---|---|---|---|
| 0 | 0 | 0 | 0 | 0 | 1 | 0 | 0 | 0 | 0 | 0 | 0 | 0 |
| 0 | 0 | 0 | 1 | 0 | 1 | 0 | 0 | 0 | 0 | 0 | 0 | 0 |
| 0 | 0 | 1 | 0 | 0 | 0 | 1 | 0 | 0 | 0 | 0 | 0 | 0 |
| 0 | 0 | 1 | 1 | 0 | 0 | 1 | 0 | 0 | 0 | 0 | 0 | 0 |
| 0 | 1 | 0 | 0 | 0 | 1 | 0 | 0 | 0 | 0 | 0 | 0 | 0 |
| 0 | 1 | 0 | 1 | 0 | 1 | 0 | 0 | 0 | 0 | 0 | 0 | 0 |
| 0 | 1 | 1 | 0 | 0 | 0 | 0 | 0 | 1 | 0 | 0 | 0 | 0 |
| 0 | 1 | 1 | 1 | 0 | 0 | 0 | 0 | 1 | 0 | 0 | 0 | 0 |
| 1 | 0 | 0 | 0 | 0 | 0 | 0 | 1 | 0 | 0 | 0 | 0 | 0 |
| 1 | 0 | 0 | 1 | 0 | 0 | 0 | 1 | 0 | 0 | 0 | 0 | 0 |
| 1 | 0 | 1 | 0 | 0 | 0 | 1 | 0 | 0 | 0 | 0 | 0 | 0 |
| 1 | 0 | 1 | 1 | 0 | 0 | 1 | 0 | 0 | 0 | 0 | 0 | 0 |
| 1 | 1 | 0 | 0 | 0 | 0 | 0 | 1 | 0 | 0 | 0 | 0 | 0 |
| 1 | 1 | 0 | 1 | 0 | 0 | 0 | 1 | 0 | 0 | 0 | 0 | 0 |
| 1 | 1 | 1 | 0 | 0 | 0 | 0 | 0 | 1 | 0 | 0 | 0 | 0 |
| 1 | 1 | 1 | 1 | 0 | 0 | 0 | 0 | 1 | 0 | 0 | 0 | 0 |
| 0 | 0 | 0 | 0 | 1 | 0 | 0 | 0 | 0 | 1 | 0 | 0 | 0 |
| 0 | 0 | 0 | 1 | 1 | 0 | 0 | 0 | 0 | 0 | 0 | 1 | 0 |
| 0 | 0 | 1 | 0 | 1 | 0 | 0 | 0 | 0 | 1 | 0 | 0 | 0 |
| 0 | 0 | 1 | 1 | 1 | 0 | 0 | 0 | 0 | 0 | 0 | 1 | 0 |
| 0 | 1 | 0 | 0 | 1 | 0 | 0 | 0 | 0 | 1 | 0 | 0 | 0 |
| 0 | 1 | 0 | 1 | 1 | 0 | 0 | 0 | 0 | 0 | 0 | 0 | 1 |
| 0 | 1 | 1 | 0 | 1 | 0 | 0 | 0 | 0 | 0 | 1 | 0 | 0 |
| 0 | 1 | 1 | 1 | 1 | 0 | 0 | 0 | 0 | 0 | 0 | 0 | 1 |
| 1 | 0 | 0 | 0 | 1 | 0 | 0 | 0 | 0 | 0 | 0 | 1 | 0 |
| 1 | 0 | 0 | 1 | 1 | 0 | 0 | 0 | 0 | 0 | 0 | 1 | 0 |
| 1 | 0 | 1 | 0 | 1 | 0 | 0 | 0 | 0 | 0 | 0 | 1 | 0 |
| 1 | 0 | 1 | 1 | 1 | 0 | 0 | 0 | 0 | 0 | 0 | 1 | 0 |
| 1 | 1 | 0 | 0 | 1 | 0 | 0 | 0 | 0 | 0 | 0 | 1 | 0 |
| 1 | 1 | 0 | 1 | 1 | 0 | 0 | 0 | 0 | 0 | 0 | 0 | 1 |
| 1 | 1 | 1 | 0 | 1 | 0 | 0 | 0 | 0 | 0 | 0 | 1 | 0 |
| 1 | 1 | 1 | 1 | 1 | 0 | 0 | 0 | 0 | 0 | 0 | 0 | 1 |

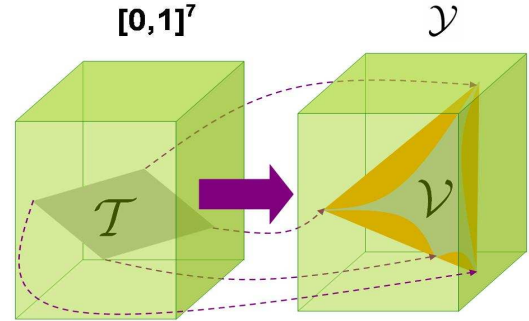

Figure 3: Transformation of entire polytope

## 4 Bounding Causal Effects

In this section we again consider the model from §3, where $C \perp\!\!\!\perp A \,|\, (B, U)$, but now we make the additional assumption that $A \perp\!\!\!\perp U$. These are the only two non-causal assumptions that we make. Here we consider causal effects and must therefore use a DAG to incorporate causal assumptions or a causal DAG (Pearl, 2000). When we say we fix the value of $A$ we mean actively hold the value of $A$ at a certain value and not passively observe that $A$ takes the value. Since the graph is acyclic then intervention on $A$ and $B$ cannot simultaneously affect each others distribution. In order to represent interventions we use the notation $(\cdot \,||\, \cdot)$, first used in Lauritzen (2001), for *intervention conditioning* and the usual $(\cdot \,|\, \cdot)$ for *observation conditioning*. The '$||$' is equivalent to the '$do$' notation, first used in Goldszmidt and Pearl (1992). An augmented directed graph (Dawid, 2002) for the model is given in Fig.4, which is Fig.1 augmented with the intervention variable $F_B$. The intervention variable represents a strategy that we set and by including it on the graph we can derive the implications of our chosen decision on the distribution of the random variables.

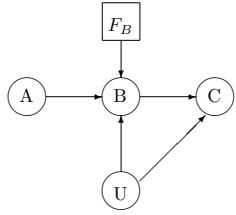

Figure 4: With causal assumptions

The absence of edges between $F_B$ and $C$, $A$ and $U$ represents non-trivial causal assumptions. The node $F_B$ takes the values '*idle*', 0 or 1. If $F_B = idle$ then $B$ takes a random value. However, if $F_B$ is either 0 or 1 then $B = F_B$. Therefore $C \perp\!\!\!\perp B \,|\, (F_B = B, U)$. This is the only conditional independence relation that is not represented on the graph. All other relations can be derived in the same way as for the purely probabilistic DAGs, using the 'moralisation criterion' (Lauritzen et al., 1990) or 'd-separation' (Verma and Pearl, 1988). Since $C \perp\!\!\!\perp B \,|\, (F_B = B, U)$, $C \perp\!\!\!\perp F_B \,|\, (B, U)$ and $U \perp\!\!\!\perp F_B$, from Fig.4,

$$
\begin{aligned}
P(C \,||\, B) &= P(C \,|\, F_B = B) \\
&= \sum_u P(C \,|\, U, F_B = B) P(U \,|\, F_B) \\
&= \sum_u P(C \,|\, U, F_B = B, B) P(U \,|\, F_B) \\
&= \sum_u P(C \,|\, U, B) P(U)
\end{aligned}
$$

Also
$$
\begin{aligned}
P(C \,|\, A) &= \sum_u P(C \,|\, A, U) P(U) \\
P(B \,|\, A) &= \sum_u P(B \,|\, A, U) P(U).
\end{aligned}
$$

Let $\vec{v} = (\gamma_{01}^*, \gamma_{11}^*, \gamma_{02}^*, \gamma_{12}^*, \theta_{01}^*, \theta_{11}^*, \theta_{02}^*, \theta_{12}^*, \alpha^*)$ and $\vec{u} = (\gamma_{01}, \gamma_{11}, \gamma_{02}, \gamma_{12}, \theta_{01}, \theta_{11}, \theta_{02}, \theta_{12}, \alpha)$, where

$$
\begin{aligned}
\gamma_{ca}^* &= P(C \,|\, A, U) \\
\gamma_{ca} &= P(C \,|\, A) \\
\theta_{ba}^* &= P(B \,|\, A, U) \\
\theta_{ba} &= P(B \,|\, A) \\
\alpha^* &= P(C = 1 \,|\, B = 1, U) - P(C = 1 \,|\, B = 0, U) \\
\alpha &= P(C = 1 \,||\, B = 1) - P(C = 1 \,||\, B = 0).
\end{aligned}
$$

Therefore any possible $\vec{u}$ lies in the convex hull of the set of possible $\vec{v}$ since $P(U) \geq 0$ and $\sum P(U) = 1$. The mapping $\Xi(\cdot)$ can be expressed as follows

$$
\begin{aligned}
\gamma_{01}^* &= (1-\eta_0)(1-\delta_1) + (1-\eta_1)\delta_1 \\
\gamma_{11}^* &= \eta_0(1-\delta_1) + \eta_1\delta_1 \\
\gamma_{02}^* &= (1-\eta_0)(1-\delta_2) + (1-\eta_1)\delta_2 \\
\gamma_{12}^* &= \eta_0(1-\delta_2) + \eta_1\delta_2 \\
\theta_{01}^* &= 1 - \delta_1 \\
\theta_{11}^* &= \delta_1 \\
\theta_{02}^* &= 1 - \delta_2 \\
\theta_{12}^* &= \delta_2 \\
\alpha^* &= \eta_1 - \eta_0.
\end{aligned}
$$

We then apply the same technique as §3 to find the constraints involving the $\gamma$'s, $\theta$'s and $\alpha$. If we leave out the $\alpha^*$ and $\alpha$ components from $\vec{v}$ and $\vec{u}$ respectively we find constraints on $(\vec{\gamma}, \vec{\theta})$. The transformation of the extreme vertices in both cases can be seen in Fig.5. We only consider half of the set of extreme vertices since $P(A \,|\, U)$ is irrelevant for the transformation. In such a case $\dim(\mathcal{V}) = 4$ but $\dim(\mathcal{T}) = 5$. This is because here $\Xi$ is a many to one mapping.

| $\eta_0$ | $\eta_1$ | $\delta_1$ | $\delta_2$ | $\psi$ | | $\gamma_{01}^*$ | $\gamma_{11}^*$ | $\gamma_{02}^*$ | $\gamma_{12}^*$ | $\theta_{01}^*$ | $\theta_{11}^*$ | $\theta_{02}^*$ | $\theta_{12}^*$ | $\alpha^*$ |
|---|---|---|---|---|---|---|---|---|---|---|---|---|---|---|
| 0 | 0 | 0 | 0 | 0 | | 1 | 0 | 1 | 0 | 1 | 0 | 1 | 0 | 0 |
| 0 | 0 | 0 | 1 | 0 | | 1 | 0 | 1 | 0 | 1 | 0 | 0 | 1 | 0 |
| 0 | 0 | 1 | 0 | 0 | | 1 | 0 | 1 | 0 | 0 | 1 | 1 | 0 | 0 |
| 0 | 0 | 1 | 1 | 0 | | 1 | 0 | 1 | 0 | 0 | 1 | 0 | 1 | 0 |
| 0 | 1 | 0 | 0 | 0 | | 1 | 0 | 1 | 0 | 1 | 0 | 1 | 0 | 1 |
| 0 | 1 | 0 | 1 | 0 | | 1 | 0 | 0 | 1 | 1 | 0 | 0 | 1 | 1 |
| 0 | 1 | 1 | 0 | 0 | | 0 | 1 | 1 | 0 | 0 | 1 | 1 | 0 | 1 |
| 0 | 1 | 1 | 1 | 0 | $\rightarrow$ | 0 | 1 | 0 | 1 | 0 | 1 | 0 | 1 | 1 |
| 1 | 0 | 0 | 0 | 0 | | 0 | 1 | 0 | 1 | 1 | 0 | 1 | 0 | -1 |
| 1 | 0 | 0 | 1 | 0 | | 0 | 1 | 1 | 0 | 1 | 0 | 0 | 1 | -1 |
| 1 | 0 | 1 | 0 | 0 | | 1 | 0 | 0 | 1 | 0 | 1 | 1 | 0 | -1 |
| 1 | 0 | 1 | 1 | 0 | | 1 | 0 | 1 | 0 | 0 | 1 | 0 | 1 | -1 |
| 1 | 1 | 0 | 0 | 0 | | 0 | 1 | 0 | 1 | 1 | 0 | 1 | 0 | 0 |
| 1 | 1 | 0 | 1 | 0 | | 0 | 1 | 0 | 1 | 1 | 0 | 0 | 1 | 0 |
| 1 | 1 | 1 | 0 | 0 | | 1 | 0 | 0 | 1 | 0 | 1 | 1 | 0 | 0 |
| 1 | 1 | 1 | 1 | 0 | | 0 | 1 | 0 | 1 | 0 | 1 | 0 | 1 | 0 |

Figure 5: Transformation of polytope

If we omit the causal effects from our analysis we obtain the following constraints

$$
\begin{aligned}
\theta_{01} + \theta_{02} &\geq \gamma_{01} - \gamma_{02} \\
\theta_{01} + \theta_{02} &\geq \gamma_{02} - \gamma_{01} \\
\theta_{11} + \theta_{12} &\geq \gamma_{01} - \gamma_{02} \\
\theta_{11} + \theta_{12} &\geq \gamma_{02} - \gamma_{01},
\end{aligned} \quad (4)
$$

or

$$
2 - |\gamma_{01} - \gamma_{02}| \geq \theta_{01} + \theta_{02} \geq |\gamma_{01} - \gamma_{02}|, \quad (5)
$$

in addition to the trivial constraints $\vec{\gamma} \geq 0$, $\vec{\theta} \geq 0$ and $\sum_c \gamma_{ca} = \sum_b \theta_{ba} = 1 \,\forall\, a$. It is important to note that these inequalities cannot be used to determine which distributions actually fit the model but can be used as a test to determine which distributions do not fit the model. Therefore Eq.(5) can be considered as a version of the 'instrumental inequality' (Pearl 1995), given in Eq.(1), when we only have $(C \,|\, A)$ and $(B \,|\, A)$ data.

If we now include the causal effects we obtain various constraints. The constraints involving observables only are the same as Eq.(5), in addition to the trivial constraints, and the constraints involving $\alpha$ are

$$
\alpha \geq \max \left\{ \begin{array}{c} 2\gamma_{01} - \gamma_{02} + 2\theta_{01} - 3 \\ \gamma_{01} + \theta_{01} - 2 \\ \gamma_{02} + \theta_{02} - 2 \\ -\gamma_{01} + 2\gamma_{02} + 2\theta_{02} - 3 \\ -\gamma_{01} + \gamma_{02} - \theta_{01} + \theta_{02} - 1 \\ -\gamma_{01} - \theta_{01} \\ -\gamma_{02} - \theta_{02} \\ \gamma_{01} - 2\gamma_{02} - 2\theta_{02} \\ -2\gamma_{01} + \gamma_{02} - 2\theta_{01} \\ \gamma_{01} - \gamma_{02} + \theta_{01} - \theta_{02} - 1 \end{array} \right\}
$$

$$\alpha \leq \min \left\{ \begin{array}{c} -2\gamma_{01} + \gamma_{02} + 2\theta_{01} + 1 \\ \gamma_{01} - 2\gamma_{02} + 2\theta_{02} + 1 \\ 2\gamma_{01} - \gamma_{02} - 2\theta_{01} + 2 \\ -\gamma_{01} + 2\gamma_{02} - 2\theta_{02} + 2 \\ \gamma_{01} - \gamma_{02} - \theta_{01} + \theta_{02} + 1 \\ -\gamma_{02} + \theta_{02} + 1 \\ \gamma_{01} - \theta_{01} + 1 \\ \gamma_{02} - \theta_{02} + 1 \\ -\gamma_{01} + \theta_{01} + 1 \\ -\gamma_{01} + \gamma_{02} + \theta_{01} - \theta_{02} + 1 \end{array} \right\}$$

If we have observational data the $\alpha$'s are not obtainable. Therefore, without experimental data, we cannot use all of the constraints as a test to determine whether a distribution is invalid under the model. Instead we use Eq.(5) to test whether a distribution fits and if it is not invalid then we use the constraints involving $\alpha$ to bound $\alpha$. Eq.(5) has to be checked separately, ahead of bounding. This is because the bounds do not automatically satisfy Eq.(5) once they are non-empty, just like the cases considered in Balke and Pearl (1997) and Dawid (2003). Provided that Eq.(5) holds, we assume that the model is valid and therefore the inequalities involving the $\alpha$'s will be satisfied. Therefore the validity of our bounds relies on this assumption and since we cannot observe $\alpha$ there is no way to check it. In §5 we present two applications of the bounds.

## 5  Tri-variate Data

Using a similar bounding technique for the model in Fig.4, we can find bounds in terms of $P(C, B \mid A)$. We can transform $\vec{\tau}$ to a vector with components $P(C, B \mid A, U)$ and $\alpha^*$. Since

$$P(C, B \mid A) = \sum_u P(C, B \mid A, U) P(U),$$

then the vector of $P(C, B \mid A)$ and $\alpha$ lies in the convex hull of the set of vectors with components $P(C, B \mid A, U)$ and $\alpha^*$. The bounds and constraints produced by this method with only the assumption $C \perp\!\!\!\perp A \mid (B, U)$ are $\vec{\eta} \geq 0$, $\sum_c \sum_b \zeta_{cb.a} = 1 \; \forall \, a$,

$$\begin{array}{c} \zeta_{00.1} + \zeta_{10.2} \leq 1 \\ \zeta_{10.1} + \zeta_{00.2} \leq 1 \\ \zeta_{11.1} + \zeta_{01.2} \leq 1 \\ \zeta_{01.1} + \zeta_{11.2} \leq 1, \end{array} \quad (6)$$

and

$$\alpha \geq \max \left\{ \begin{array}{c} \zeta_{00.1} + \zeta_{11.2} - 1 \\ \zeta_{11.1} + \zeta_{00.2} - 1 \\ -\zeta_{01.1} - \zeta_{10.1} + \zeta_{11.1} - \zeta_{10.2} - \zeta_{11.2} \\ -\zeta_{10.1} - \zeta_{11.1} - \zeta_{01.2} - \zeta_{10.2} + \zeta_{11.2} \\ -\zeta_{01.1} - \zeta_{10.1} \\ -\zeta_{01.2} - \zeta_{10.2} \\ -\zeta_{00.1} - \zeta_{01.1} + \zeta_{00.2} - \zeta_{01.2} - \zeta_{10.2} \\ \zeta_{00.1} - \zeta_{01.1} - \zeta_{10.1} - \zeta_{00.2} - \zeta_{01.2} \end{array} \right\}$$

Table 1: Probability distribution derived from Lipid Research Clinics Coronary Primary Prevention Trial (1984).

| $a$ | $\zeta_{00.a}$ | $\zeta_{01.a}$ | $\zeta_{10.a}$ | $\zeta_{11.a}$ |
|---|---|---|---|---|
| 1 | 0.919 | 0 | 0.081 | 0 |
| 2 | 0.315 | 0.139 | 0.073 | 0.473 |

| $a$ | $\theta_{0a}$ | $\theta_{1a}$ | $a$ | $\gamma_{0a}$ | $\gamma_{1a}$ |
|---|---|---|---|---|---|
| 1 | 1 | 0 | 1 | 0.919 | 0.081 |
| 2 | 0.388 | 0.612 | 2 | 0.454 | 0.546 |

| $b$ | $\phi_{0b}$ | $\phi_{1b}$ |
|---|---|---|
| 0 | 0.623 | 0.077 |
| 1 | 0.068 | 0.232 |

$$\alpha \leq \min \left\{ \begin{array}{c} 1 - \zeta_{10.1} - \zeta_{01.2} \\ 1 - \zeta_{01.1} - \zeta_{10.2} \\ \zeta_{00.1} - \zeta_{01.1} + \zeta_{11.1} + \zeta_{00.2} + \zeta_{01.2} \\ \zeta_{00.1} + \zeta_{01.1} - \zeta_{01.2} + \zeta_{00.2} + \zeta_{11.2} \\ \zeta_{00.1} + \zeta_{11.1} \\ \zeta_{00.2} + \zeta_{11.2} \\ \zeta_{10.1} + \zeta_{11.1} + \zeta_{00.2} + \zeta_{11.2} - \zeta_{10.2} \\ \zeta_{00.1} - \zeta_{10.1} + \zeta_{11.1} + \zeta_{10.2} + \zeta_{11.2} \end{array} \right\},$$

where $\zeta_{cb.a} = P(C, B \mid A)$. These are the same as those in Balke and Pearl (1997) and Dawid (2003) for the model in which $A$ is an instrument for the effect of $B$ on $C$. The method employed here is exactly that used in Dawid (2003).

We consider the Lipid Research Clinics Coronary Primary Prevention Trial data (Lipid Research Clinic Program, 1984), which was analyzed by Efron and Feldman (1991) and Balke and Pearl (1997). Subjects were randomized into two groups, 172 men were given the placebo and 165 were given the treatment, and the subjects' cholesterol levels were measured. There was partial compliance of the subjects with the treatment assigned. The probabilities estimated from the data are given in Table 1, where $\phi_{cb} = P(C, B)$. We assume the estimates are from large samples and are therefore considered as estimates from the entire population. The bounds from §4 and this section are applied for the causal effect of treatment on response and a comparison given below. The $\phi$ data are not used here.

**Coronary Primary Prevention Trial**

BP (1997) and D(2003), $(C, B \mid A)$ data:

$$\alpha \geq \max \left\{ \begin{array}{c} 0.392, -0.685, -0.627, 0.18, \\ -0.081, -0.212, -0.816, 0.384 \end{array} \right\}$$

$$\alpha \leq \min \left\{ \begin{array}{c} 0.78, 0.927, 1.373, 1.568, \\ 0.919, 0.788, 0.796, 1.384 \end{array} \right\}$$

$$\Rightarrow \quad 0.392 \leq \alpha \leq 0.780$$

Table 2: Probability Distribution derived from vitamin A data of Sommer and Zeger (1991).

| $a$ | $\zeta_{00.a}$ | $\zeta_{01.a}$ | $\zeta_{10.a}$ | $\zeta_{11.a}$ |
|---|---|---|---|---|
| 1 | 0.0064 | 0 | 0.9936 | 0 |
| 2 | 0.0028 | 0.0010 | 0.1972 | 0.7990 |

| $a$ | $\theta_{0a}$ | $\theta_{1a}$ |
|---|---|---|
| 1 | 1 | 0 |
| 2 | 0.2 | 0.8 |

| $a$ | $\gamma_{0a}$ | $\gamma_{1a}$ |
|---|---|---|
| 1 | 0.0064 | 0.9936 |
| 2 | 0.0038 | 0.9962 |

| $b$ | $\phi_{0b}$ | $\phi_{1b}$ |
|---|---|---|
| 0 | 0.0046 | 0.5882 |
| 1 | 0.0005 | 0.4067 |

$(C \,|\, A)$ and $(B \,|\, A)$ data:

$$\alpha \geq \max \left\{ \begin{array}{c} 0.384, -0.081, -1.158, -2.235, \\ -2.077, -1.919, -0.842, \\ -0.765, -3.384, 0.077 \end{array} \right\}$$

$$\alpha \leq \min \left\{ \begin{array}{c} 1.616, 1.787, 1.384, 1.213, \\ 0.853, 0.934, 0.919, \\ 1.066, 1.081, 1.147 \end{array} \right\}$$

$$\Rightarrow \quad 0.384 \leq \alpha \leq 0.853$$

Remarkably, the width of the bounds for the bivariate and trivariate data are not much different. Therefore we are still able to make useful inference when given the less informative bivariate data. Another example of partial compliance is the study of Vitamin A supplementation in northern Sumatra described by Sommer and Zeger (1991). The study consisted of children in 450 villages, 221 villages were assigned to the control group and 229 to the treatment group and the estimated probabilities are given in Table 2. Those assigned to the control group were not given a placebo because of government policy. The causal effects were also analyzed by Balke and Pearl (1997).

**Vitamin A Supplementation**

BP (1997) and D(2003), $(C, B \,|\, A)$ data:

$$\alpha \geq \max \left\{ \begin{array}{c} -0.1946, -0.9972, -1.9898, \\ -0.3928, -0.9936, -0.1982, \\ -0.2018, -0.991 \end{array} \right\}$$

$$\alpha \leq \min \left\{ \begin{array}{c} 0.0054, 0.8028, 0.0102, \\ 0.8072, 0.0064, 0.8018, \\ 1.5982, 0.009 \end{array} \right\}$$

$$\Rightarrow \quad -0.1946 \leq \alpha \leq 0.0054$$

$(C \,|\, A)$ and $(B \,|\, A)$ data:

$$\alpha \geq \max \left\{ \begin{array}{c} -0.991, -0.9936, -1.7962, \\ -2.5988, -1.8026, -1.0064, \\ -0.2038, -0.4012, -2.009, -0.1974 \end{array} \right\}$$

$$\alpha \leq \min \left\{ \begin{array}{c} 2.991, 1.3988, 0.009, \\ 1.6012, 0.2026, 1.1962, \\ 0.0064, 0.8038, 1.9936, 1.7974 \end{array} \right\}$$

$$\Rightarrow \quad -0.1974 \leq \alpha \leq 0.0064$$

Here again the bounds are very similar. The similarity of the bounds when given the bivariate and trivariate data for these two examples may not necessarily hold in general. For each of the examples the data from all of the tables were derived from the same study but, ignoring sampling uncertainty, the bounds still apply if each individual table was obtained from a different study. Also in both cases it was checked separately that the data satisfied Eqs.(5) and (6) before the bounds were calculated. The sampling uncertainty in data was ignored in the above analyses, but by using techniques similar to those in Cheng and Small (2006) it would be possible to quantify the probability that the causal effect is within specific bounds.

## 6 Other Constraints

In this section we consider two examples with the same model of Fig.4 in which $C \perp\!\!\!\perp A \,|\, (B, U)$ and $A \perp\!\!\!\perp U$, but bound various causal effects given different sets of data. In §6.2 we investigate the effect of extra data on the bounds in §4.

### 6.1 Example 1

Consider a study in which we want to make inference about the causal effect of $A$ on $C$ and only have data on the effect of $A$ on $B$. We can easily use the same technique as §4 to obtain the required bounds, by considering the columns for $\theta^*$ in Fig.5 and one for $\beta^*$ where

$$\beta^* = P(C = 1 \,|\, A = 2, U) - P(C = 1 \,|\, A = 1, U).$$

However this is not necessary in this case since the same bounds on $\beta$ can be easily derived from Eq.(4), where

$$\beta = P(C = 1 \,||\, A = 2) - P(C = 1 \,||\, A = 1).$$

This is because $P(C \,||\, A) = P(C \,|\, A)$ for the model under consideration. Therefore $\beta = \gamma_{12} - \gamma_{11}$, so the non-trivial bounds

$$\max \left\{ \begin{array}{c} -\theta_{01} - \theta_{02} \\ -\theta_{11} - \theta_{12} \end{array} \right\} \leq \beta \leq \min \left\{ \begin{array}{c} \theta_{01} + \theta_{02} \\ \theta_{11} + \theta_{12} \end{array} \right\}.$$

directly follow from Eq.(4).

### 6.2 Example 2

In section §4 we derived the bounds for the causal effect of $B$ on $C$ when given two sets of bivariate data. Consider a study in which we now have extra data available, i.e. we now have $(C \,|\, A)$, $(B \,|\, A)$ and $(C, B)$ data. Therefore bounds are needed which can incorporate this additional information and narrow the possible range of the causal effect. Here also it is in principle not necessary that each pairwise set of data come from the same study, although this ensures consistency of the pairwise distributions. The transformation of the extreme vertices are given below.

| $\gamma^*_{01}$ | $\gamma^*_{11}$ | $\gamma^*_{02}$ | $\gamma^*_{12}$ | $\theta^*_{01}$ | $\theta^*_{11}$ | $\theta^*_{02}$ | $\theta^*_{12}$ | $\phi^*_{00}$ | $\phi^*_{01}$ | $\phi^*_{10}$ | $\phi^*_{11}$ | $\alpha^*$ |
|---|---|---|---|---|---|---|---|---|---|---|---|---|
| 1 | 0 | 1 | 0 | 1 | 0 | 1 | 0 | 1 | 0 | 0 | 0 | 0 |
| 1 | 0 | 1 | 0 | 0 | 0 | 1 | 0 | 1 | 0 | 0 | 0 | 0 |
| 1 | 0 | 1 | 0 | 0 | 1 | 1 | 0 | 0 | 1 | 0 | 0 | 0 |
| 1 | 0 | 1 | 0 | 0 | 0 | 1 | 0 | 1 | 0 | 0 | 0 | 0 |
| 1 | 0 | 1 | 0 | 1 | 0 | 1 | 0 | 0 | 1 | 0 | 0 | 1 |
| 1 | 0 | 0 | 1 | 1 | 0 | 1 | 0 | 1 | 0 | 0 | 0 | 1 |
| 0 | 1 | 1 | 0 | 0 | 1 | 1 | 0 | 0 | 0 | 1 | 1 | 1 |
| 0 | 1 | 0 | 1 | 0 | 1 | 0 | 1 | 0 | 0 | 0 | 1 | 1 |
| 0 | 1 | 0 | 1 | 1 | 0 | 1 | 0 | 0 | 0 | 1 | 0 | -1 |
| 0 | 1 | 1 | 0 | 1 | 0 | 1 | 0 | 0 | 0 | 1 | 0 | -1 |
| 1 | 0 | 0 | 1 | 0 | 1 | 1 | 0 | 0 | 1 | 0 | 0 | -1 |
| 1 | 0 | 1 | 0 | 0 | 1 | 0 | 1 | 0 | 1 | 0 | 0 | -1 |
| 0 | 1 | 0 | 1 | 1 | 0 | 0 | 1 | 0 | 0 | 1 | 0 | 0 |
| 0 | 1 | 0 | 1 | 1 | 0 | 0 | 1 | 0 | 0 | 1 | 0 | 0 |
| 0 | 1 | 0 | 1 | 0 | 1 | 1 | 0 | 0 | 0 | 0 | 1 | 0 |
| 0 | 1 | 0 | 1 | 0 | 1 | 0 | 1 | 0 | 0 | 1 | 0 | 0 |
| 1 | 0 | 1 | 0 | 1 | 0 | 1 | 0 | 1 | 0 | 0 | 0 | 0 |
| 1 | 0 | 1 | 0 | 1 | 0 | 0 | 1 | 1 | 0 | 0 | 0 | 0 |
| 1 | 0 | 1 | 0 | 0 | 1 | 1 | 0 | 1 | 0 | 0 | 0 | 0 |
| 1 | 0 | 1 | 0 | 0 | 1 | 0 | 1 | 1 | 0 | 0 | 0 | 0 |
| 1 | 0 | 1 | 0 | 1 | 0 | 1 | 0 | 1 | 0 | 0 | 0 | 1 |
| 1 | 0 | 1 | 0 | 1 | 0 | 0 | 1 | 0 | 1 | 0 | 0 | 1 |
| 0 | 1 | 1 | 0 | 0 | 1 | 1 | 0 | 0 | 0 | 0 | 1 | 1 |
| 0 | 1 | 0 | 1 | 0 | 1 | 1 | 0 | 0 | 0 | 0 | 1 | 1 |
| 0 | 1 | 0 | 1 | 1 | 0 | 1 | 0 | 0 | 0 | 1 | 0 | -1 |
| 0 | 1 | 1 | 0 | 1 | 0 | 1 | 0 | 1 | 0 | 0 | 0 | -1 |
| 1 | 0 | 0 | 1 | 1 | 0 | 0 | 1 | 0 | 0 | 1 | 0 | -1 |
| 1 | 0 | 1 | 0 | 0 | 1 | 0 | 1 | 1 | 0 | 0 | 0 | -1 |
| 0 | 1 | 0 | 1 | 1 | 0 | 1 | 0 | 1 | 0 | 0 | 1 | 0 |
| 0 | 1 | 0 | 1 | 1 | 0 | 0 | 1 | 0 | 0 | 0 | 1 | 0 |
| 0 | 1 | 0 | 1 | 0 | 1 | 1 | 0 | 0 | 0 | 1 | 0 | 0 |
| 0 | 1 | 0 | 1 | 0 | 1 | 0 | 1 | 0 | 0 | 1 | 0 | 0 |

where $\phi^*_{cb} = P(C = c, B = b \,|\, U)$. The resulting constraints are $\vec{\theta} \geq 0$, $\vec{\gamma} \geq 0$, $\vec{\phi} \geq 0$, $\sum \theta = \sum \gamma = \sum \phi = 1$, 56 inequalities involving observable variables and 37 inequalities each for the upper and lower bounds. Since the number of constraints are so large we do not reproduce them here. However when the bounds were applied to the data from Table 1 and 2 the following were obtained. The 56 involving observable variables were checked ahead of bounding to ensure the model was not invalid.

**Coronary Primary Prevention Trial**

$(C \,|\, A)$, $(B \,|\, A)$ and $(C, B)$ data:

$$0.388 \leq \alpha \leq 0.851$$

**Vitamin A Supplementation**

$(C \,|\, A)$, $(B \,|\, A)$ and $(C, B)$ data:

$$-0.1974 \leq \alpha \leq 0.0059$$

These bounds do not vastly improve those with two sets of bivariate data and are wider than the trivariate data bounds. Therefore not much information is gained in this case with the extra data but in other situations the bounds may be narrowed significantly. We did not consider the case of having the extra data in the form $P(C \,|\, B)$ because a vector having components $P(C \,|\, B)$ would not lie in the convex hull of the set of vectors with components $P(C \,|\, B, U)$ since

$$P(C \,|\, B) \neq \sum_u P(C \,|\, B, U) P(U).$$

## 7 Conclusion and Discussion

The general method of bounding causal effects can also be applied to other graphical models but the transformed function must be finite and the observable quantities must be in the convex hull of the transformed components. Causal bounds can be used to make inference about causal effects when only limited data is available and can collapse to point estimates, be vague or even be trivial. This does not discredit the techniques discussed because they were meant to be used in scenarios where the ability to make any inference at all is very useful, especially with the possibility of the bounds collapsing to a point estimate. It has been shown that the bounds exist by direct use of the properties of probability distributions, without relying on the additional assumptions associated with potential outcomes (Dawid, 2000). Our current research also involves making assumptions and assessing their effects on the width of the bounds.

In our approach, conditional independence relations constrain the observational distribution and, together with causal assumptions in the form of conditional independence relations involving interventions, constrain causal effects. Both types of assumptions can be represented in an augmented DAG. The method readily produces tests such as the instrumental inequality, which can be used to invalidate a model, and bounds on unobservable causal effects. However, we assume that the model holds if the constraints involving available data hold. As with any assumption, it can be strengthened by expert knowledge.

Using the bounds derived, inference can be made when only bivariate data are available in situations involving instrumental variables, with the added benefit that pairwise data from more than one study can be used. The availability of extra data can narrow the width of the estimated causal bounds but its effect on their accuracy is unknown. It will be very useful to quantify the uncertainty associated with the bounds. Then we might be able to determine whether the trivariate, 2 sets of bivariate or 3 sets of bivariate data bounds have very different uncertainties associated with them despite the similarity of the estimates of the bounds. Bounds have only been developed for the binary case but for more complicated variables the computations are possible but become quite long and might need to be implemented via computer programming. However, the technique of bounding causal effects is a valuable alternative when the data required for more precise analyses is unavailable.

### Acknowledgements

Thanks to Steffen Lauritzen for first suggesting this research and for his insightful supervision of it. I am also grateful to Phil Dawid for valuable discussions which improved the paper and Mathias Drton for suggesting the use of Polymake. This research was funded by a scholarship from the Government of Trinidad and Tobago.

## A  Appendix: Properties of Convex Hull

Here we prove that $\mathcal{H} = \hat{\mathcal{H}}$, following Dawid (2003). Since $\hat{\mathcal{V}} \subseteq \mathcal{V}$ and $\hat{\mathcal{H}}$ is the minimal convex set containing $\hat{\mathcal{V}}$ then $\hat{\mathcal{H}} \subseteq \mathcal{H}$.
Let $m(\vec{v})$ be an affine function of $\vec{v}$ for $\vec{v} \in \hat{\mathcal{V}}$. Consider the inequality (i.e. closed half space in $[0,1]^8$) $m(\vec{v}) \geq 0$ or $m(\Xi(\vec{\tau})) \geq 0$ for $\vec{\tau} \in \hat{\mathcal{T}}$. Since $m(\Xi(\tau))$ is a monotonic function of any component of $\vec{\tau}$ when the other three are fixed, then the minimum of $m(\Xi(\vec{\tau}))$ over $\mathcal{T}$ is attained when $\vec{\tau} \in \hat{\mathcal{T}}$. Therefore

$$m(\Xi(\vec{\tau})) \geq 0 \text{ for } \vec{\tau} \in \hat{\mathcal{T}} \Rightarrow m(\Xi(\vec{\tau})) \geq 0 \text{ for } \vec{\tau} \in \mathcal{T}$$

This means that any half space containing $\hat{\mathcal{V}}$ also contains $\mathcal{V}$. Since $\hat{\mathcal{H}}$ is the intersection of all half spaces containing $\hat{\mathcal{V}}$ then $\mathcal{V} \subseteq \hat{\mathcal{H}}$. Since $\hat{\mathcal{H}}$ is convex and $\mathcal{H}$ is the minimal convex set containing $\mathcal{V}$ then $\mathcal{H} \subseteq \hat{\mathcal{H}}$. Therefore $\mathcal{H} = \hat{\mathcal{H}}$.